\documentclass[10pt, conference, letterpaper]{IEEEtran}

\usepackage{subfiles}
\usepackage{xspace}
\usepackage{graphicx}

\usepackage{enumitem} %controls indentation of items in an itemized list 

\newcommand{\subparagraph}{}
\usepackage[utf8]{inputenc}
\usepackage[english]{babel}
\newcommand{\papertitle}{WAKU-RLN-RELAY\xspace}
\newcommand{\waku}{WAKU\xspace}
\newcommand{\wakurelay}{WAKU-RELAY\xspace}
\newcommand{\wakufilter}{12/WAKU2-FILTER\xspace}
\newcommand{\wakustore}{13/WAKU2-STORE\xspace}
\newcommand{\treeroot}{\tau\xspace}

\usepackage{tikz} % for IEEE copyright
\usepackage{textcomp}
\usepackage{hyperref}
\newcommand\copyrighttext{%
  \footnotesize \textcopyright 
  20xx IEEE. Personal use of this material is permitted. Permission from IEEE must be obtained for all other uses, in any current or future media, including reprinting/republishing this material for advertising or promotional purposes, creating new collective works, for resale or redistribution to servers or lists, or reuse of any copyrighted component of this work in other works.
  }
\newcommand\copyrightnotice{%
\begin{tikzpicture}[remember picture,overlay]
\node[anchor=south,yshift=10pt] at (current page.south) {\fbox{\parbox{\dimexpr\textwidth-\fboxsep-\fboxrule\relax}{\copyrighttext}}};
\end{tikzpicture}%
}
\IEEEoverridecommandlockouts % to display thanks https://tex.stackexchange.com/questions/53546/thanks-wont-appear-in-ieeetran

\begin{document}
\title{\papertitle: Privacy-Preserving Peer-to-Peer Economic Spam Protection}

\author{\IEEEauthorblockN{Sanaz Taheri-Boshrooyeh\IEEEauthorrefmark{1}\IEEEauthorrefmark{2},
Oskar Thor\'en\IEEEauthorrefmark{1}\IEEEauthorrefmark{2}, 
Barry Whitehat\IEEEauthorrefmark{3},
Wei Jie Koh\IEEEauthorrefmark{4},
Onur Kilic\IEEEauthorrefmark{5}, and
Kobi Gurkan\IEEEauthorrefmark{6}}
\IEEEauthorblockA{
\IEEEauthorrefmark{1}Vac Research and Development,
\IEEEauthorrefmark{2}Status Research and Development, Singapore,
\IEEEauthorrefmark{3}Unaffiliated,\\
\IEEEauthorrefmark{4}Independent,
\IEEEauthorrefmark{5}Unaffiliated,
\IEEEauthorrefmark{6}cLabs 
}
\IEEEauthorblockA{
% \IEEEauthorrefmark{1}
sanaz@status.im,
% \IEEEauthorrefmark{2}
oskar@status.im,
% \IEEEauthorrefmark{3}
barrywhitehat@protonmail.com,\\
% \IEEEauthorrefmark{4}
contact@kohweijie.com,
% \IEEEauthorrefmark{5}
onurkilic@protonmail.com,
% \IEEEauthorrefmark{6}
me@kobi.one
}
}
\maketitle
\copyrightnotice

\begin{abstract}
In this paper, we propose \papertitle as a spam-protected gossip-based routing protocol that can run in heterogeneous networks. It features a privacy-preserving peer-to-peer (p2p) economic spam protection mechanism.
\papertitle addresses the performance and privacy issues of the state-of-the-art p2p spam prevention techniques including peer scoring utilized by libp2p, and proof-of-work used by e.g., Whisper, the p2p messaging layer of Ethereum. In \papertitle, spam protection works by limiting the messaging rate of each network participant. Rate violation is disincentivized since it results in financial punishment where the punishment is cryptographically guaranteed. Peers who identify spammers are also rewarded.
To enforce the rate limit, we adopt the suggested framework of Semaphore and its extended version, however, we modify that framework to properly address the unique requirements of a network of p2p resource-restricted users. The current work dives into the end-to-end integration of Semaphore into \papertitle, the modifications required to make it suitable for resource-limited users, and the open problems and future research directions. We also provide a proof-of-concept open-source implementation of \papertitle, and its specifications together with a rough performance evaluation. 

\end{abstract}

\section{Introduction}
\waku \cite{vacrfc} is a family of peer-to-peer (p2p) protocols for anonymous and privacy-preserving communication. It is designed to be able to run in resource-restricted environments e.g., limited computational power, bandwidth, and storage space.  Being p2p means that \waku relies on no central server. Instead, peers collaboratively deliver messages in the network.  \wakurelay constitutes the transport layer of \waku and aims at being privacy-preserving in which no one knows the owner and the receiver of a message except the two ends of the communication. Many of the design choices in this layer are centered around the anonymity requirement.  \wakurelay follows a publisher-subscriber messaging model and adopts a gossip-based routing protocol. It is a thin layer over the libp2p GossipSub \cite{vyzovitis2020gossipsub} routing protocol. In a pubsub messaging model, peers congregate around topics they are interested in and can send messages to topics. Each message gets delivered to all peers subscribed to the topic. Each peer has a constant number of direct connections/neighbors. To publish a message, the author forwards its message to a subset of neighbors. The neighbors proceed similarly till the message gets propagated in the network of the subscribed peers.  The adopted gossip-based routing allows concealing the message receiver hence greatly benefits receiver anonymity \cite{unger2015sok}.  

In addition to \wakurelay, \waku features other types of protocols for running in resource-restricted environments. Among which are the request/response protocols of  \wakufilter \cite{vacrfc} which is a lightweight version of \wakurelay for devices with limited bandwidth, and \wakustore \cite{vacrfc} by which resourceful peers can persist and offer historical messages to the querying nodes. Details of these protocols are out of the scope of this paper and can be found in the Waku RFCs \cite{vacrfc}. 

\begin{figure*}[h]
\centering
\includegraphics[scale=0.17, trim={2cm 15cm 3cm 3cm}]{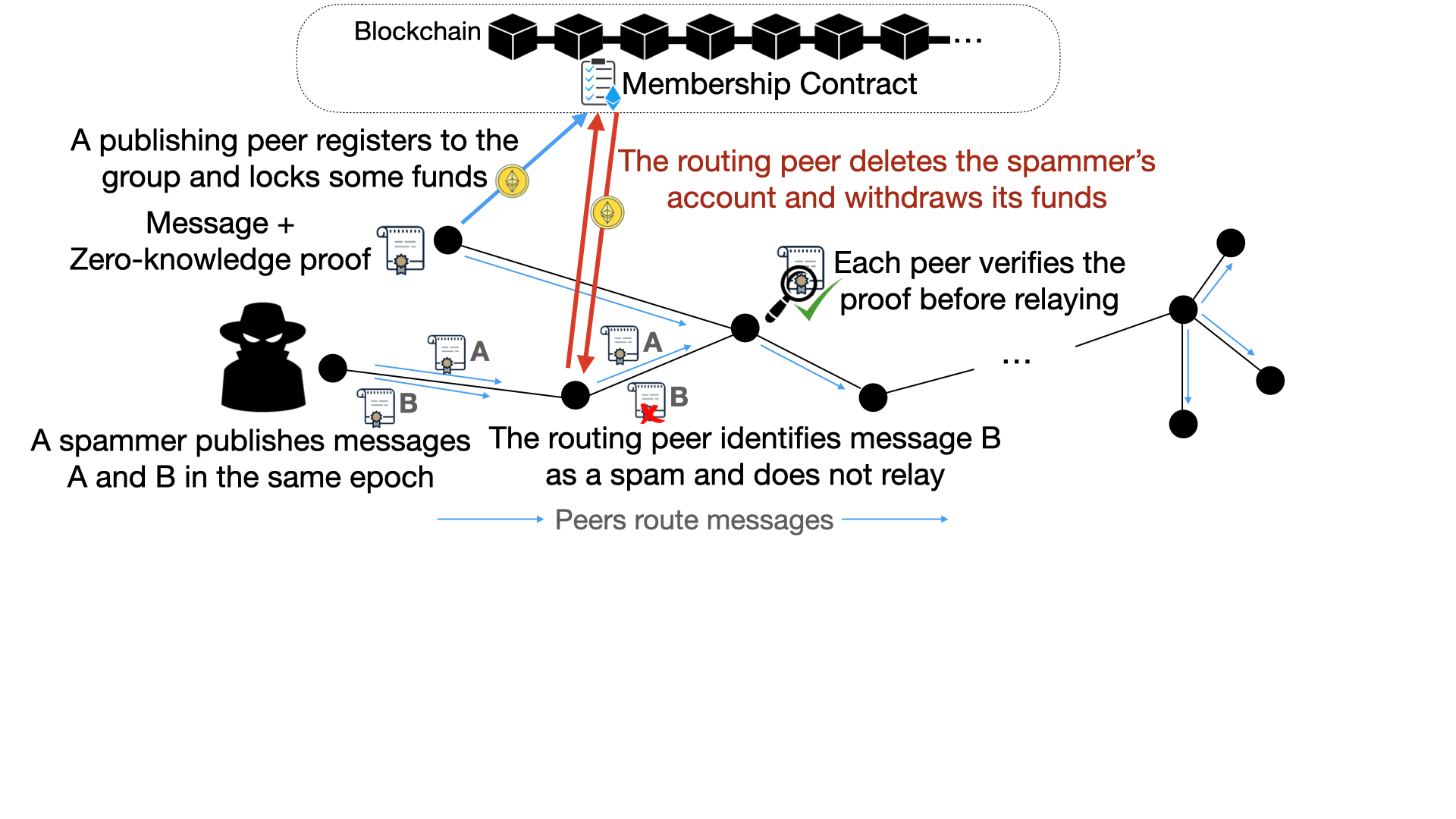} 
    \caption{An overview of privacy-preserving p2p economic spam protection in \papertitle protocol.}
    \label{fig:rln-relay-overview}
\end{figure*} 

As an open messaging network, \wakurelay is prone to spam messages. A spammer usually indicates an entity that uses the messaging system to send an \textit{unsolicited} message (spam) to \textit{large numbers of recipients}. However, in \wakurelay with an open and gossip-based structure, spam messages not only affect the recipients but also all the other peers involved in the routing process where they have to spend their resources e.g., computational power, bandwidth, and storage capacity on processing spam messages. As such, we define a spammer as an entity that uses the messaging system to publish a large number of messages in a short amount of time, in other words, has a high messaging rate. We define messages issued in this way as spam regardless of the intention of the spammer, the content of the message, and the number of intended recipients.

The state-of-the-art p2p spam protection techniques for messaging systems i.e.,  \textit{Proof of Work} (PoW) \cite{dwork1992pricing}  deployed by e.g, Whisper \cite{whisper} and \textit{Peer scoring} \cite{vyzovitis2020gossipsub} method adopted by libp2p have their own shortcomings. The PoW technique imposes a high computational cost for messaging hence devices with limited resources won't be able to participate and benefit from the messaging system. The peer scoring method is prone to censorship and is also subject to inexpensive attacks where the spammer can send bulk messages by deploying millions of bots. 
The spam protection methods deployed by centralized messaging platforms e.g., WhatsApp exhibit privacy issues where they usually ask users to disclose and commit to some piece of personally identifiable information e.g., phone number and email address at the registration time. In addition to this, the central provider is aware of messages owned and received by a particular user which is against privacy.

% The centralized spam protection methods exhibit privacy issues where they usually ask users to disclose and commit to some piece of personally identifiable information e.g., phone number and email address at the registration time. In addition to this, the central provider is aware of messages owned and received by a particular user which is against privacy.

The economic-incentive spam protection mechanism of  \papertitle aims at coping with the aforementioned issues where
\begin{enumerate}
    \item It suits \textbf{p2p} systems and does not rely on any central entity.
    \item It is \textbf{efficient} in terms of computation, storage, memory, and bandwidth usage, hence suitable for a network of \textit{heterogeneous} peers with limited resources.
    % i.e., computational power, bandwidth, and storage space. 
    \item It respects users' \textbf{privacy} unlike reputation-based and centralized methods. It also allows global identification and removal of spammers while not requiring personally identifiable information from the participants.
    \item It deploys \textbf{economic-incentives} to contain spammers' activity. Namely, there is a financial sacrifice for those who want to spam the system.  Additionally, there is a financial reward to monitor the network for spam and remove the offending users.

\end{enumerate}

At a high level, \papertitle guarantees no one can publish more than one message per \textit{epoch} i.e., a time interval, without being financially charged. The spammer gets removed from the network and will not be able to publish further messages. The \textit{epoch} length is a system parameter that should be configured to meet the desired messaging rate and application throughput. For example, while an epoch duration of $1$ second i.e., a messaging rate of $1$ per second might be acceptable for a chat application, might be too low for communication among Ethereum network validators.

\section{Preliminaries: Rate Limiting Nullifier}
\subsection{Semaphore}
Semaphore \cite{gurkan2020community} is a zero-knowledge signaling framework on Ethereum. It allows a set of users to broadcast arbitrary signals (where the signal is any value like a string, vote, etc.) while proving they are among a group of authorized users without disclosing their identities. Use-cases of Semaphore are anonymous authentication and private voting. It also utilizes external nullifiers to prevent double-signaling. An external nullifier can be seen as a voting booth where each user can only cast one vote \cite{semaphore.appliedzkp}. Casting a second vote for the same booth will be rejected. Similarly, each signal is bound to an external nullifier, and each group member is allowed to signal only once for that external nullifier. While nullifiers can limit the signal rate per user, there is no way to identify and remove users who violate this rate limit. In an attempt to address this shortcoming, an extended version of this method, using Shamir Secret Sharing (SSS) \cite{shamir1979share} has been proposed called Rate-Limiting Nullifier or RLN for short. In RLN \cite{semaphore-slash}, if a user attempts more than one signal for the same external nullifier, it reveals its identity/private key by which it has registered to the group. As such, the identified identity key can be removed from the group and the user won't be able to signal anymore. The removed user will be also financially punished. The user initially deposits some funds when joining the group, and the fund will be rewarded to anyone who identifies double signaling of that user.

\subsection{RLN: Semaphore with Shamir Secret Sharing} \label{sec:semaphore}
In this section, we present an overview of the extended version of Semaphore that utilizes Shamir secret sharing \cite{shamir1979share} to enable the identification of spammers. 

Each member has a private key $sk$ and a public key $pk=H(sk)$ where $H$ is a Cryptographic hash function. These are also called the identity key and the identity commitment, respectively. The list of group members is stored in an \textit{Identity Commitment Tree} which is a Merkle tree \cite{merkle1987digital} whose leaves are members' public keys. The root of the tree is denoted by $\treeroot$. The membership of a user in the identity commitment tree is proven by providing a branch of the tree that connects the root to that leaf corresponding to the user's $pk$. This branch is called \textit{authentication path }and we denote it by $auth$. The tree is stored on a smart contract deployed on the Ethereum blockchain. Each user additionally needs to deposit some funds in the contract at the time of registration. 

Signaling is bound to a publicly known \textit{external nullifier} denoted by $\emptyset$.
When publishing signal $m$, a group member utilizes (2,n)-Shamir secret sharing to derive a share $(x,y)$ of its private identity key $sk$  where $x=H(m)$ and $y=sk + H(sk, \emptyset)*x$. $(x,y)$  will be published alongside with the signal $m$. 

The publishing user also calculates an \textit{internal nullifier} $\phi$ as $\phi= H(H(sk, \emptyset)))$ and publishes it together with the signal $m$. 
Finally, a user needs to  prove in a zero-knowledge manner that
% \todo{page 9 of Semaphore for formal statements}
\begin{enumerate}
    \item its secret key $sk$ belongs to the identity commitment tree namely i.e., provides proof of membership. 
    \item $(x,y)$ is a valid share of its identity key.
    \item the internal nullifier $\phi$ is correctly calculated.
\end{enumerate}

The above items are proven through \textit{zkSNARKs} i.e., \textit{a zero-knowledge Succinct Non-interactive Argument of Knowledge} proof system where the circuit represents the aforementioned constraints. The public inputs to the zero-knowledge proof system are the message hash, the external nullifier $\emptyset$, internal nullifier $\phi$, the share of identity secret key  $(x,y)$ and the tree root $\treeroot$ whereas the private inputs provided by the message owner are the identity secret key $sk$, the index of $pk$ in the tree, and the authentication path $auth$.
In \papertitle, we utilize Groth16 \cite{groth2016size} for the proof system. The parameter generation can be done through a multi-party setup similar to \cite{bowe2017scalable, trustedsetupceremony, powersoftau-software, powersoftau-output}.

The user submits its signal $m$,  identity key share $(x,y)$, and internal nullifiers $\phi$ to the contract to be stored and accessible by the rest of the group members. The contract maintains the state of all the group signals together with all the metadata that can be used to identify double signaling.
If a user attempts two different signals for the same external nullifier $\emptyset$,  their internal nullifiers will collide, which signifies a double signaling attempt. Furthermore, the two shares of the user's identity key can be used to reconstruct its $sk$ and remove the user from the group. The $sk$ reconstruction stems from the fact that each line (polynomial of degree 1) can be uniquely reconstructed by using two distinct points on it. In the case of double signaling, the spammer reveals two distinct shares $(x, y)$ and $(x’, y’)$ on line $y=sk + H(sk, \emptyset)*x$, which enables the reconstruction of the line and its evaluation at $x=0$ which is $sk$. 
The user who reconstructs the $sk$ also gets rewarded by a portion of the slashed user's stake. This can be done by communicating the recovered $sk$ to the contract.

\section{\papertitle Construction }
In this section, we describe the flow of the economic spam detection mechanism in \papertitle as also depicted in Figure \ref{fig:rln-relay-overview}. 

\subsection{Overview of \papertitle vs Semaphore}
\papertitle adopts the extended variant of Semaphore i.e., RLN which provides the additional capability of identifying spammers. However, further adjustments are required to make it fit a p2p routing system.
Following is the list of these adjustments that mainly affect the state of the contract.
\begin{enumerate}[leftmargin=*]
\item In \papertitle the state of the contract keeps a simple ordered list of users' identity commitments (instead of Merkle tree) and the Merkle tree is kept off-chain by individual peers. This is in contrast to the original setting of Semaphore where the contract holds the entire commitment tree. The reason for this design shift is to mitigate the significant computational cost/ gas consumption associated with member insertion and deletion which is logarithmic in the number of registered members. While the cost associated with insertion in Semaphore could be amortized by using batch insertion, this solution does not apply to the deletion since deletion affects random leaves of the tree which cannot be necessarily batched together. \papertitle mitigates this problem as insertion and deletion modify a single item of the list.
\item  In \papertitle, users' messages, and their metadata are not stored in the contract which is in contrast to Semaphore. In \papertitle messages are stored off-chain and are distributed through the \wakurelay routing protocol. It has the benefits of 1) having higher message propagation speed and  1) being more economic (messaging is for free) as opposed to the on-chain message store. In the on-chain message storage of  Semaphore, published messages will not be visible until blocks containing those message transactions get mined. This results in an unnecessary and undesirable delay which is not acceptable for messaging systems with e.g., 1.1 million messages per second\footnote{https://www.oberlo.ca/blog/whatsapp-statistics}. In an attempt to address this issue, \papertitle decouples the message propagation and storage from the contract state and provides a p2p and off-chain medium for message transportation i.e., \wakurelay \cite{vacrfc} and storage i.e., \wakustore \cite{vacrfc}.
    The off-chain storage, adopted by \papertitle,  has another advantage of being more economic. That is it saves the financial cost (related to the gas consumption) associated with message insertion into the contract state. This cost for one-time messaging scenarios like voting systems can be tolerable, however, it is far from practical in a messaging application where there are millions of messages transmitted per second. 
\end{enumerate}

Due to these adjustments, sending messages in \papertitle is for free i.e., does not need gas consumption. Furthermore, message transmission is not affected by the underlying blockchain and its consensus layer. Such separation allows the utilization of various optimization techniques on the message transmission delay which would be otherwise impossible due to reliance on the blockchain.

\subsection{Setup and Registration}
\begin{figure}[h]
\centering
\includegraphics[width=0.5\textwidth, trim = {0 2cm 0 0 }]{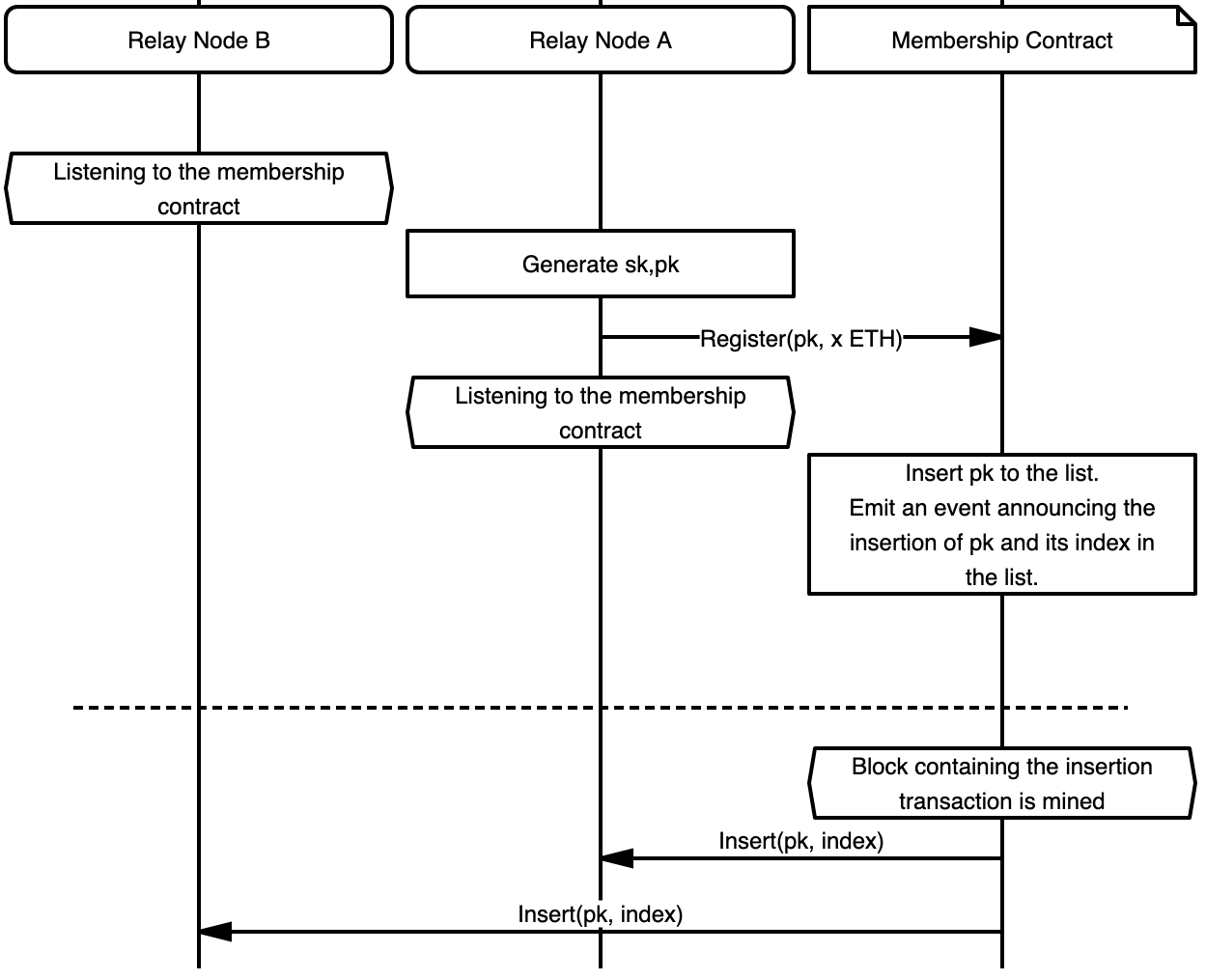} 
    \caption{Overview of registration process.}
    \label{fig:registration}
\end{figure}

A peer willing to publish a message is required to register. Registration is moderated through a smart contract deployed on the public Ethereum blockchain. The state of the contract contains the list of registered members' public keys i.e., identity commitment keys. An overview of registration is illustrated in Figure \ref{fig:registration}.

For the registration, a peer creates its identity private key $sk$ and its commitment $pk=H(sk)$ and sends a transaction to the contract to register its identity commitment $pk$ in the group. The transaction also transfers the $v$ amount of Ether to the contract. This amount is deposited on the contract to prevent spam activity.

\subsection{Maintaining the identity commitment tree}
The construction and maintenance of the identity commitment tree, unlike the original proposal of Semaphore \cite{gurkan2020community}, is delegated to the peers. As we previously mentioned, the reason is that the cost associated with member deletion and insertion is high and unreasonable.
As such, each peer needs to build the tree locally and listen to the contract's events i.e., peer insertion and deletion, and update its tree accordingly.
Publishing peers must always stay in sync with the latest state of the group. Otherwise,  by making proof of membership to an old version of the membership tree they can risk exposing the index of their public key in the tree hence compromising their anonymity.

\subsection{External nullifier}
For the external nullifier, we define $epoch$ which is some unit of time elapsed since the Unix epoch i.e., January 1, 1970  (midnight UTC/GMT).  The $epoch$ length is denoted by $T$ and its value is up to the application. It is measured as $UnixTime/ T$ where $UnixTime$ indicates Unix epoch time e.g., if $UnixTime$ is $1644810116$ seconds and the epoch length is set to $30$ seconds, then $epoch=\lceil 1644810116s/30s\rceil= 54827003$. Peers locally keep track of the current $epoch$ and can publish one message per $epoch$.

\subsection{Publishing}

\begin{figure*}
\centering
\includegraphics[scale=0.21, trim={0 1cm 0 0}]{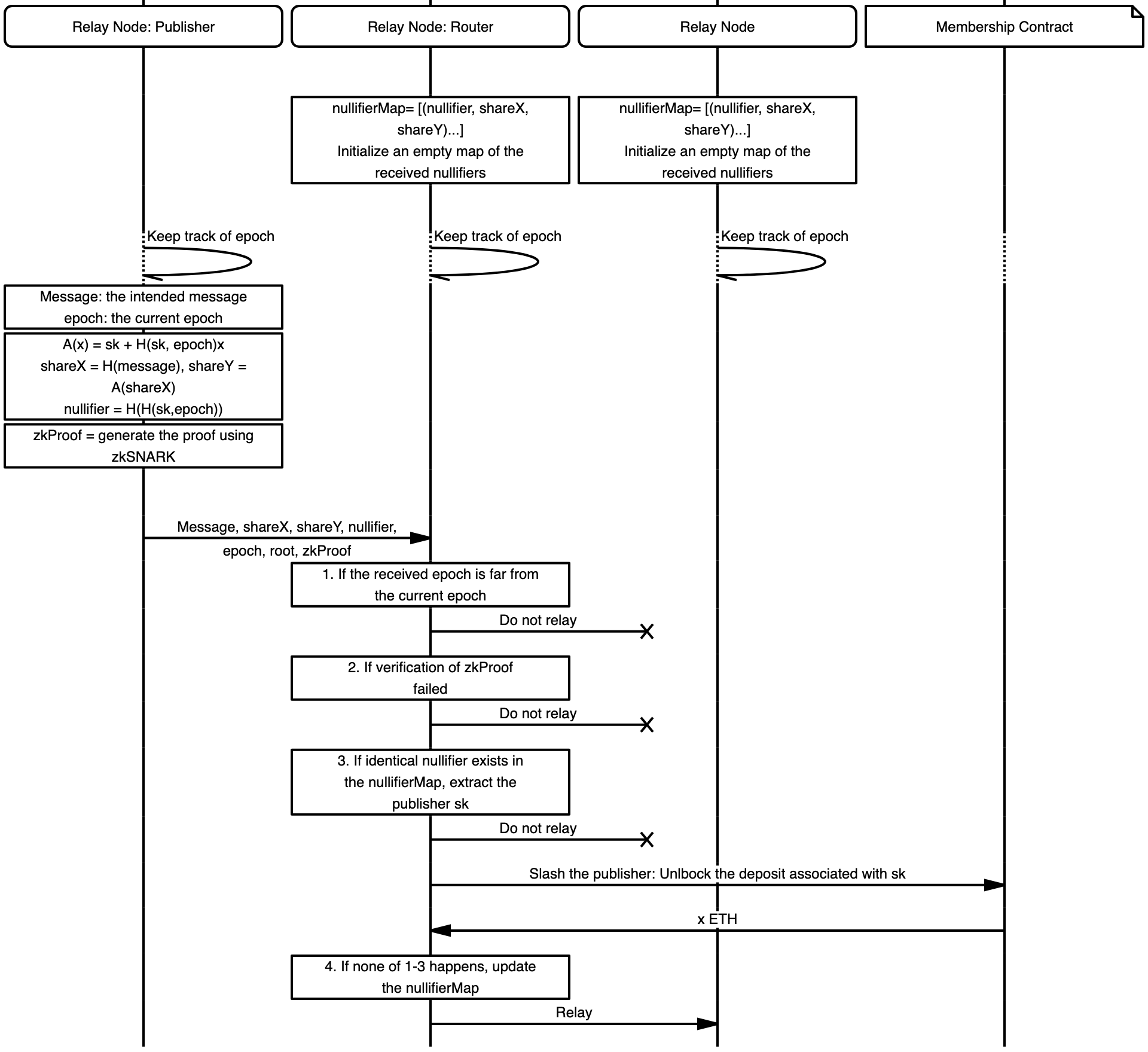}
    \caption{Overview of message publishing, routing and slashing.}
    \label{fig:outline}
\end{figure*}

Each peer is allowed to send one message $m$ per $epoch$ without being slashed/financially punished. The financial punishment is that the fund deposited by the spammer is rewarded to the peer reporting spammer's activity. To prove that the peer is not a spammer and has not violated the messaging rate for the current epoch, the peer is required to generate some metadata and send them alongside the message $m$. The metadata includes all the public inputs to the zkSNARKs as explained in \ref{sec:semaphore} i.e., a share of the peer's identity secret key i.e., $(x,y)$, the internal nullifier $\phi$, and the external nullifier $\emptyset = epoch$ and the root of identity commitment tree $\treeroot$ and zkSNARKs proof $\pi$. The peer then sends the message bundle $(m, (x,y), \phi, epoch, \treeroot, \pi)$ to its direct connections as instructed by the routing algorithm i.e., \wakurelay. An overview of the publishing procedure is provided in Figure \ref{fig:outline}.

\subsection{Routing and Slashing} \label{sec:routing_slashing}
Peers follow the regular routing protocol of \wakurelay and in addition check the metadata of each relayed message to identify and spot spam messages and slash spammers. The routing procedure is depicted in Figure \ref{fig:outline}. 

Upon the receipt of a message $(m, (x,y), \phi, epoch, \treeroot, \pi)$, the routing peer needs to decide whether to route it or not. The decision relies on the following factors:
\begin{enumerate}
    \item  If the $epoch$ value attached to the message has more than $Thr$  gap with the routing peer's current epoch, the message is considered invalid and must be dropped. This is to prevent a newly registered peer from spamming the system by messaging for all the past epochs. $Thr$ is a system parameter.
    \item The message must contain valid proof $\pi$ that gets verified by the routing peer. 
    \item The messaging rate is not violated.
\end{enumerate}
If the preceding checks are passed successfully, then the message is relayed. In case of invalid proof, the message is dropped. If spamming is detected, the publishing peer gets slashed and the slasher gets rewarded.

To identify spam messages, each routing peer keeps a local record of the identity key share $(x,y)$ and the internal nullifier $\phi$ of all of its valid incoming message bundles for the past $Thr$ epochs. This list is called a nullifier map. The routing peer checks every new message against this list to spot spam messages.  This list does not have to capture the entire history because any incoming message whose epoch is older than the last $Thr$ epochs is discarded by default (as explained in item 1 of section \ref{sec:routing_slashing}). The routing peer utilizes this list to locally identify spam messages and spammers as follows.

\begin{enumerate}
    \item The routing peer first verifies the $\pi$ and discards the message if not verified.
    \item It checks for the presence of another past message with an identical internal nullifier i.e., $\phi$. If nothing is found, then the message gets relayed, otherwise:
    \begin{enumerate}
        \item  If the identity share i.e., $(x',y')$ of the older message is different from the incoming message share i.e., $(x,y) \neq (x',y')$, then slashing takes place 
        \item If $(x,y) = (x',y')$, then the message is a duplicate and should be discarded.
    \end{enumerate}
\end{enumerate}

\textbf{Race condition:} There is a possibility of a race condition when slashing an identified spammer. The race happens when a peer submits the identified identity key of the spammer to the contract in plain format and at the same time, another peer observes and copies the same information to reclaim the spammer's fund. This yields a race issue. To avoid this, the commit and reveal technique can be utilized, i.e., the routing peer sends a commitment to the spammer's identity key (instead of the plain identity key) to the contract so that no one else can replicate and reclaim it. Later it opens the commitment and reveals its knowledge of the spammer's identity key.

\textbf{Maximum Epoch Gap:} As mentioned earlier, the gap between the epoch observed by the routing peer and the one attached to the incoming message should not exceed a threshold denoted by $Thr$. $Thr$  can be measured based on  1) the network transmission delay which is the maximum time that it takes for a message to be fully disseminated in the network 2) the clock asynchrony i.e., the maximum difference between the Unix epoch time perceived by the network peers which can be due to clock drifts. With a reasonable approximation of the preceding values, one can set  $ Thr = \lceil \frac{Network Delay + Clock Asynchrony}{ T}\rceil$ where $T$ is the length of the epoch in seconds (the $Network Delay$ and $Clock Asynchrony$ are also in seconds). By this formulation, $Thr$ measures the maximum number of epochs that can elapse since a message gets routed from its origin to all the other peers in the network.

\section{Security and Performance Analysis}

\textbf{Security:} \papertitle achieves global spam protection while preserving user anonymity, namely 1) Peers do not have to disclose any piece of personally identifiable information in any phase i.e., neither in the registration nor the messaging phase 2) Peers can prove that they have not exceeded the messaging rate in a zero-knowledge manner and without leaving any trace to their identity public keys.  Our solution also protects network resources against spammers as the spam messages are dropped immediately and not propagated. Malicious participants that may attempt to send messages with invalid proofs to exhaust the resources of the network will also fail because the effect of their attack is 1) limited to their direct connections and will not impact the entire network, since such messages are not relayed further due to invalid proof 2) is easily addressable by leveraging peer scoring mechanism \cite{vyzovitis2020gossipsub}. Sybil attack is also mitigated by making registration expensive. Economic incentives are guaranteed cryptographically via secret sharing. 

\textbf{Performance:} 
The proof of concept implementation as well as specifications of \papertitle are available in \cite{nim-waku} and \cite{vacrfc}, respectively. We utilize  RLN library \cite{rlnlib} for our implementation. Generating membership proof to a group size of  $2^{32}\approx 4$  billion peers takes $\approx0.5$s on an iPhone 8 \cite{rlnlib}. Proof verification run time is constant and takes $\approx30$ms. Each peer persists a $32$B public and secret keys and a prover key with $\approx3.89$MB in size \cite{rlnlib}.
Storage of a membership tree with depth $20$ takes up $67$MB from each peer (this can be optimized to $0.128$KB using the proposal of \cite{storage-efficient}).

\subsection{Future Work} \label{sec:FutureWork}
\papertitle is currently a Proof of Concept (POC), and its design and development are in progress. Below, we shed light on some of our future work.

\textbf{Evaluating Merkle tree computation overhead:} We would like to evaluate the running time associated with the Merkle tree operations. Indeed, the need to locally store Merkle tree on each peer was one of the unknowns discovered during this POC and yet the concrete benchmarking result in this regard is not available. 

\textbf{Enhancing performance by off-chain solutions:} Another possible improvement is to replace the membership contract with a distributed group management scheme e.g., through distributed hash tables. This is to address possible performance issues that the interaction with the public Ethereum blockchain may cause.  For example, the registration transactions are subject to delay as they have to be mined before being visible in the state of the membership contract. This means peers have to wait for some time before being able to publish any message. The same issue exists for slashing and other smart contract-related functions. The use of state channels and optimistic Rollups are other ways to overcome transaction delays and achieve better performance.

\textbf{Lowering the storage overhead per peer}: Currently, peers need to maintain the entire tree locally which incurs a storage overhead linear in the size of the group. This overhead can be tackled using a hybrid architecture where peers with adequate storage capacity retain the tree and supply the necessary information to the resource-limited peers upon request. Alternatively, the storage efficient method of \cite{storage-efficient}  can be utilized where peers maintain a partial view of the tree yet can construct and update the tree root and their authentication paths based on the group updates.  Keeping the partial view lowers the storage complexity to $O(log(N))$ where $N$ is the size of the group. In this solution, the authentication path of the slashed keys can become available assuming that peers, at the registration time,  1) provide their authenticated path encrypted under their secret identity key and 2) prove the former in a zero-knowledge manner. The use of Verkle tree \cite{kuszmaul2019verkle} or polynomial commitments \cite{kate2010polynomial} are  other paths to follow. 

\textbf{Cost-effective member insertion and deletion}:
Depending on Ethereum gas costs, the cost associated with membership is $40k$  gas which translates to more than $20$ USD (at the time of writing of this paper). We aim at finding a more cost-effective approach.  For example, by using batch insertion and deletion, the cost can be reduced to $20k$ gas. Alternatively, layer two solutions or more cost-efficient blockchains with support for smart contracts can be utilized.

\subsection{Open Problems}
Below is the list of open problems for which no immediate solution is known hence demanding more long-term research.

\textbf{Exceeding the messaging rate via multiple registrations}:
While the economic-incentive solution has an economic incentive to discourage spamming,  there is still the possibility of expensive attack(s) for exceeding the messaging rate. An attacker pays for multiple e.g., $k$ registrations, and uses its aggregate quota for messaging i.e., $k$ messages per epoch. Such attacks can be mitigated by increasing the entry barrier via a higher membership fee. Following this argument, the high fee associated with the membership, which is discussed in section \ref{sec:FutureWork}, can indeed be beneficial for spam prevention.

\noindent\textbf{Escaping punishment by early withdrawal}: 
A spammer can escape from getting slashed by withdrawing its fund from the contract before its spam activity gets caught. While this means the attacker burns its initial membership fund (the fee paid to register its key to the group), it allows saving the other part of the fund that can be otherwise rewarded to slashers.

\section*{Acknowledgment}
We acknowledge Corey Petty\footnote{Corey@status.im} for his valuable comments on the paper.

\bibliographystyle{IEEEtran}
\bibliography{refs}

\begin{thebibliography}{10}
\providecommand{\url}[1]{#1}
\csname url@samestyle\endcsname
\providecommand{\newblock}{\relax}
\providecommand{\bibinfo}[2]{#2}
\providecommand{\BIBentrySTDinterwordspacing}{\spaceskip=0pt\relax}
\providecommand{\BIBentryALTinterwordstretchfactor}{4}
\providecommand{\BIBentryALTinterwordspacing}{\spaceskip=\fontdimen2\font plus
\BIBentryALTinterwordstretchfactor\fontdimen3\font minus
  \fontdimen4\font\relax}
\providecommand{\BIBforeignlanguage}[2]{{%
\expandafter\ifx\csname l@#1\endcsname\relax
\typeout{** WARNING: IEEEtran.bst: No hyphenation pattern has been}%
\typeout{** loaded for the language `#1'. Using the pattern for}%
\typeout{** the default language instead.}%
\else
\language=\csname l@#1\endcsname
\fi
#2}}
\providecommand{\BIBdecl}{\relax}
\BIBdecl

\bibitem{vacrfc}
``Specifications of waku protocols stack,'' \url{ https://rfc.vac.dev/},
  accessed: 2022-03.

\bibitem{vyzovitis2020gossipsub}
D.~Vyzovitis, Y.~Napora, D.~McCormick, D.~Dias, and Y.~Psaras, ``Gossipsub:
  Attack-resilient message propagation in the filecoin and eth2. 0 networks,''
  \emph{arXiv preprint arXiv:2007.02754}, 2020.

\bibitem{unger2015sok}
N.~Unger, S.~Dechand, J.~Bonneau, S.~Fahl, H.~Perl, I.~Goldberg, and M.~Smith,
  ``Sok: secure messaging,'' in \emph{2015 IEEE Symposium on Security and
  Privacy}.\hskip 1em plus 0.5em minus 0.4em\relax IEEE, 2015, pp. 232--249.

\bibitem{dwork1992pricing}
C.~Dwork and M.~Naor, ``Pricing via processing or combatting junk mail,'' in
  \emph{Annual international cryptology conference}.\hskip 1em plus 0.5em minus
  0.4em\relax Springer, 1992, pp. 139--147.

\bibitem{whisper}
``Whisper specifications,'' \url{https://eips.ethereum.org/EIPS/eip-627 },
  accessed: 2022-03.

\bibitem{gurkan2020community}
\BIBentryALTinterwordspacing
K.~Gurkan, K.~W. Jie, and B.~Whitehat, ``Community proposal: Semaphore:
  Zero-knowledge signaling on ethereum,'' 2020. [Online]. Available:
  \url{https://github.com/appliedzkp/semaphore/blob/master/spec/\\Semaphore\%20Spec.pdf}
\BIBentrySTDinterwordspacing

\bibitem{semaphore.appliedzkp}
\url{https://semaphore.appliedzkp.org/ }, accessed: 2022-03.

\bibitem{shamir1979share}
A.~Shamir, ``How to share a secret,'' \emph{Communications of the ACM},
  vol.~22, no.~11, pp. 612--613, 1979.

\bibitem{semaphore-slash}
``Rln original proposal,''
  \url{https://ethresear.ch/t/semaphore-rln-rate-limiting-nullifier-for-spam-prevention-in-anonymous-p2p-setting/5009},
  accessed: 2022-03.

\bibitem{merkle1987digital}
R.~C. Merkle, ``A digital signature based on a conventional encryption
  function,'' in \emph{Conference on the theory and application of
  cryptographic techniques}.\hskip 1em plus 0.5em minus 0.4em\relax Springer,
  1987, pp. 369--378.

\bibitem{groth2016size}
J.~Groth, ``On the size of pairing-based non-interactive arguments,'' in
  \emph{Annual international conference on the theory and applications of
  cryptographic techniques}.\hskip 1em plus 0.5em minus 0.4em\relax Springer,
  2016, pp. 305--326.

\bibitem{bowe2017scalable}
S.~Bowe, A.~Gabizon, and I.~Miers, ``Scalable multi-party computation for
  zk-snark parameters in the random beacon model,'' \emph{Cryptology ePrint
  Archive}, 2017.

\bibitem{trustedsetupceremony}
``https://medium.com/coinmonks/announcing-the-perpetual-powers-of-tau-ceremony-to-benefit-all-zk-snark-projects-c3da86af8377.''

\bibitem{powersoftau-software}
``Implementation of multi-party computation (mpc) ceremony for zk-snark
  parameters,''
  \url{https://github.com/kobigurk/phase2-bn254/tree/master/powersoftau},
  accessed: 2022-03.

\bibitem{powersoftau-output}
``First phase of a multi-party trusted setup ceremony based on zcash powers of
  tau ceremony,'' \url{ https://github.com/weijiekoh/perpetualpowersoftau},
  accessed: 2022-03.

\bibitem{nim-waku}
``Reference implementation of waku in nim,''
  \url{https://github.com/status-im/nim-waku}, accessed: 2022-03.

\bibitem{rlnlib}
``Rln library implemented in rust,'' \url{ https://github.com/kilic/rln},
  accessed: 2022-03.

\bibitem{storage-efficient}
``Storage efficient merkle tree update,'' \url{
  https://github.com/vacp2p/research/blob/master/rln-research/merkle-tree-update.md},
  accessed: 2022-03.

\bibitem{kuszmaul2019verkle}
J.~Kuszmaul, ``Verkle trees,'' \emph{Verkle Trees}, pp. 1--12, 2019.

\bibitem{kate2010polynomial}
A.~Kate, G.~M. Zaverucha, and I.~Goldberg, ``Polynomial commitments,''
  \emph{Tech. Rep}, 2010.

\end{thebibliography}

\end{document}